\documentclass[aps,prd,twocolumn,superscriptaddress,showpacs,showkeys]{revtex4}

\usepackage{epsfig,epsf}
\usepackage{amsmath}
\usepackage{amsthm}
\usepackage{amsfonts}
\usepackage{amssymb}
\usepackage{dsfont}

\usepackage{multirow}

\usepackage{slashed}

\usepackage[active]{srcltx}
\usepackage{psfrag}

\def\II{\hbox{{1}\kern-.25em\hbox{l}}}

\begin{document}

\title{Kinematic power corrections in off-forward hard reactions}

\date{\today}

\author{V.M.~Braun}
\affiliation{Institut f\"ur Theoretische Physik, Universit\"at
   Regensburg,D-93040 Regensburg, Germany}
\author{A.N.~Manashov}
\affiliation{Institut f\"ur Theoretische Physik, Universit\"at
   Regensburg,D-93040 Regensburg, Germany}
\affiliation{Department of Theoretical Physics,  St.-Petersburg State
University,
199034, St.-Petersburg, Russia}

\date{\today}

\begin{abstract}
We develop a general approach to the calculation of 
kinematic corrections $\sim t/Q^2, m^2/Q^2$ in hard processes which
involve momentum transfer from the initial to the final hadron state.
As the principal result, the complete expression is derived for the
time-ordered product of two electromagnetic currents that
includes all kinematic corrections to twist-four accuracy.
The results are immediately applicable e.g. to the studies of
deeply-virtual Compton scattering.
 \end{abstract}



\maketitle

%
%

It is generally accepted that hard exclusive scattering processes with nonzero momentum
transfer to the target can provide one with a three-dimensional picture of the proton in
longitudinal and transverse plane, encoded in generalized parton distributions
(GPDs)~\cite{Diehl:2003ny,Belitsky:2005qn}. One of the principal reactions in this context
is Compton scattering with one real and one highly-virtual photon (DVCS) which has
received a lot of attention. The QCD description of DVCS is based on the operator product
expansion (OPE) of the time-ordered product of two electromagnetic currents where the GPDs
appear as operator matrix elements and the coefficient functions can be calculated
perturbatively. In order to unravel the transverse proton structure one is interested in
particular in the dependence of the amplitude on the momentum transfer to the target
$t=(P'-P)^2$ in a reasonably broad range. Since, on the other hand, the available photon
virtualities $Q^2$ are not very large, corrections of the type $\sim t/Q^2$ which are
formally twist-four effects, can be significant and should be taken into account.

Such corrections can be called ``kinematic'' since they
only involve ratios of kinematic variables and are seemingly disconnected
from nonperturbative effects (e.g. one may consider a theoretical
limit $\Lambda_{\rm QCD}^2 \ll t \ll Q^2$). Yet the separation of
kinematic corrections $\sim t/Q^2$ from generic twist-four corrections
$\sim \Lambda_{\rm QCD}^2/Q^2$ proves to be surprisingly difficult.
The problem is well known and important for phenomenology,
as acknowledged by many 
authors~\cite{Belitsky:2005qn,Blumlein:2000cx,Radyushkin:2000ap,Belitsky:2000vx,Belitsky:2001hz,Belitsky:2010jw,Geyer:2004bx,Blumlein:2006ia,Blumlein:2008di}.

The difficulty is due to the fact that, unlike target mass corrections in
inclusive reactions \cite{Nachtmann:1973mr} which are determined
solely by the contributions of leading twist operators, the $\sim t/Q^2$ corrections
(and for spin-1/2 targets also $\sim m^2/Q^2$ corrections)
also arise from higher-twist-four operators that can be reduced
to total derivatives of the twist-two ones. Indeed,
let $\mathcal{O}_{\mu_1\ldots\mu_n}$ be a multiplicatively renormalizable
(conformal) local twist-two operator,
symmetrized and traceless over all indices. Then the operators
\begin{equation}
  \mathcal{O}_1 = \partial^2 \mathcal{O}_{\mu_1\ldots\mu_n}\,, \qquad
  \mathcal{O}_2 = \partial^{\mu_1}\mathcal{O}_{\mu_1\ldots\mu_n}
\label{eq:O1O2}
\end{equation}
are, on the one hand, twist-four, and on the other hand their matrix elements
are obviously given by the reduced leading twist matrix elements, times the
momentum transfer squared (up to, possibly, target mass corrections).
 Thus, contributions of the both operators must
be taken into account.

The problem arises because $\mathcal{O}_2$ has very peculiar properties:
divergence of a conformal operator vanishes in a free theory
(the Ferrara-Grillo-Parisi-Gatto theorem \cite{Ferrara:1972xq}).
As a consequence, using QCD equations of motion (EOM)
$\mathcal{O}_2$ can be expressed as a sum of contributions of quark-antiquark-gluon
operators. The simplest example of such relation
is~\cite{Kolesnichenko:1984dj,Braun:2004vf,Anikin:2004ja}
\begin{equation}
  \partial^\mu O_{\mu\nu} = 2i\bar q ig G_{\nu\mu}\gamma^\mu q\,,
\label{eq:puzzle}
\end{equation}
where
$
  O_{\mu\nu} = (i/2)[\bar q \gamma_\mu\!\stackrel{\leftrightarrow}{D}_\nu \!q
  + (\mu\leftrightarrow\nu)]
$
is the quark part of the energy-momentum tensor. The operator on the r.h.s.
of Eq.~(\ref{eq:puzzle}) involves the gluon field strength and, not knowing this
identity, it would be tempting to assume that
hadronic matrix elements of this operator are of the order of $\Lambda_{\rm QCD}^2$, which is wrong.
More complicated examples involving leading-twist operators with two derivatives
can be found in \cite{Balitsky:1989ry,Ball:1998ff}.

The general structure of such relations is, schematically
\begin{equation}
 (\partial\mathcal{O})_{N} = \sum_k a^{(N)}_{k}\,{G}_{Nk}\,,
\label{eq:partialO}
\end{equation}
where ${G}_{Nk}$ are twist-four quark-antiquark-gluon
(and more complicated) operators and $a^{(N)}_{k}$ the numerical coefficients.
The subscript $N$ stands for the number of
derivatives in $\mathcal{O}_N$ and the summation
goes over all contributing operators, with and without total derivatives
(so that in reality $k$ is some multi-index).
The same operators, ${G}_{Nk}$, also appear in the OPE for
the current product at the twist-four level:
\begin{equation}
 T\{j(x)j(0)\}^{t=4} = \sum_{N,k} c_{N,k}(x)\,{G}_{Nk}\,.
\label{eq:Tproduct}
\end{equation}
A separation of ``kinematic'' and ``dynamical'' contributions implies
rewriting this expansion in such a way that the contribution
of the particular combination appearing in (\ref{eq:partialO}) is
separated from the remaining twist-four contributions.
The ``kinematic'' approximation would correspond to taking into account
this term only, and neglecting contributions of ``genuine'' quark-gluon operators.

Our starting observation is that the separation of kinematic and dynamical
effects is only meaningful if they have autonomous scale dependence.
Different twist-four operators of the same dimension mix with each other
and satisfy a matrix renormalization group (RG) equation which can be solved,
at least in principle. Let $\mathcal{G}_{N,k}$ be the set of multiplicatively
renormalizable twist-four operators so that
\begin{equation}
  \mathcal{G}_{N,k} = \sum_{k'} \psi^{(N)}_{k,k'}\, {G}_{N,k'}\,.
\end{equation}
The relation (\ref{eq:partialO}) tells us that one of the solutions
of the RG equation is known {\it without the calculation}. Namely, there exists
a twist-four operator with the anomalous dimension equal to the anomalous
dimension of the leading twist operator, and Eq.~(\ref{eq:partialO})
presents the corresponding eigenvector.
(For simplicity we ignore the contributions of $\partial^2\mathcal{O}_N$
operators in this discussion; they do not pose a ``problem'' and can simply
be taken into account.)

Assume this special solution corresponds to $k=0$, so
$\mathcal{G}_{N,k=0} \equiv (\partial\mathcal{O})_{N}$ and
$\psi^{(N)}_{k=0,k'}= a_{k'}$. Inverting the
matrix of coefficients $\psi^{(N)}_{k,k'}$ we can write the expansion of an
arbitrary twist-four operator in terms of the multiplicatively renormalizable
ones
\begin{equation}
 G_{N,k} = \phi^{(N)}_{k,0} (\partial\mathcal{O})_{N} + \sum_{k'\not=0}\phi^{(N)}_{k,k'}\,
\mathcal{G}_{N,k'}\,.
\label{eq:eee}
\end{equation}
Inserting this expansion in Eq.~(\ref{eq:Tproduct}) one obtains
\begin{equation}
  T\{j(x)j(0)\}^{\rm tw-4} = \sum_{N,k} c_{N,k}(x)\phi^{(N)}_{k,0}\,(\partial\mathcal{O})_{N}
 + \ldots\,,
\label{eq:solve}
\end{equation}
where ellipses stand for the contributions of ``genuine'' twist-four
operators (with different anomalous dimensions).
The problem with this (formal) solution is that finding the coefficients
$\phi^{(N)}_{k,0}$ in general requires the knowledge of the full 
matrix $\psi^{(N)}_{k,k'}$, alias explicit solution
of the twist-four RG equations, which is not available.

Twist-four operators in QCD can be divided in two classes:
quasipartonic~\cite{Bukhvostov:1985rn}, that only involve ``plus''
components of the fields, and non-quasipartonic which also include
``minus'' light-cone projections. Our next observation is that
quasipartonic operators are irrelevant for the present discussion
since they have autonomous evolution (to the one-loop accuracy).
Hence terms in $(\partial\mathcal O)_N$ do not appear in the re-expansion
of quasipartonic operators in multiplicatively renormalizable operators, Eq.~(\ref{eq:eee}): 
the corresponding coefficients
$\phi^{(N)}_{k,0}$ vanish. As the result, the kinematic power correction
$\sim (\partial\mathcal O)_N$ is entirely due to contributions of
non-quasipartonic operators.

Renormalization of twist-four non-quasipartonic operators was
studied systematically in~\cite{Braun:2008ia,Braun:2009vc}.
The main result is that the RG
equations can be written in terms
of several $SL(2)$-invariant kernels. Using this technique,
we are able to prove that
the anomalous dimension matrix for non-quasipartonic operators is
hermitian with respect to a certain scalar product, which implies that
different eigenvectors are mutually orthogonal, i.e.
\begin{equation}
 \sum_k \mu^{(N)}_k \psi^{(N)}_{l,k} \psi^{(N)}_{m,k} \sim \delta_{l,m}\,,
\end{equation}
where $\mu^{(N)}_k$ is the corresponding (nontrivial) measure.
Using this orthogonality relation and the expression (\ref{eq:partialO})
for the relevant eigenvector, one obtains, for the non-quasipartonic
operators
\begin{equation}
 \phi^{(N)}_{k,0} = a^{(N)}_{k} ||a^{(N)}||^{-2}\,,
\label{eq:coc}
\end{equation}
where $||a^{(N)}||^2 =  \sum_k \mu^{(N)}_k (a^{(N)}_{k})^2$.
Inserting this expression in (\ref{eq:solve}) one obtains
the desired separation of kinematic effects. 

The actual derivation proves to be rather involved. It is done using the two-component 
spinor formalism in intermediate steps and requires some specific techniques
of the $SL(2)$ representation theory.
The purpose of this letter is to present our main 
result; the technical details will be given elsewhere.

We define nonlocal (light-ray) vector $O_V$ and axial-vector $O_A$  operators of the leading-twist-two
as the generating functions for local twist-two operators
\begin{align}\label{Olt}
O(z_1x,z_2x) =& \big[\bar q(z_1 x)\slashed{x}\,(\gamma_5)\, Q\,q(z_2 x )\big]_{l.t.}.
\end{align}
Here $x_\mu$ is a four-vector which is not necessarily light-like, $z_1$ and $z_2$ are
real numbers and $Q$ is the (diagonal) matrix of quark electromagnetic charges squared.
The Wilson line between the quark fields is implied. 
The leading-twist projector $[\ldots]_{l.t.}$ stands for the subtraction of traces 
of the local operators so that formally
\begin{eqnarray}
\lefteqn{
\big[\bar q(z_1 x)\slashed{x}\, Q\,q(z_2x)\big]_{l.t.} =}
\nonumber\\
&=& \sum_{N} \frac{1}{N!} x_\mu x_{\mu_1}\ldots x_{\mu_N} \Big\{\bar q(0)\gamma_\mu
[z_1\!\stackrel{\leftarrow}{D}_{\mu_1}+z_2\!\stackrel{\rightarrow}{D}_{\mu_1}]
\ldots
\nonumber\\&&{}\ldots
[z_1\!\stackrel{\leftarrow}{D}_{\mu_N}+z_2\!\stackrel{\rightarrow}{D}_{\mu_N}]Q q(0)
 -{\rm traces}\Big\}.
\end{eqnarray}
The leading-twist light-ray operators satisfy the Laplace equation
$
\partial_x^2 O(z_1x,z_2x) = 0\,.
$
Explicit form of the projector $[\ldots]_{l.t.}$  is irrelevant
for the further discussion (some useful representations can be found
e.g. in~\cite{Belitsky:2001hz,Balitsky:1987bk}).

Alternatively, one can expand a nonlocal operator in the contributions
of multiplicatively renormalizable (in one loop) conformal operators
\begin{equation}
{O}(z_1x,z_2x) =\!\sum_N\varkappa_N\, z_{12}^N\int_0^1\! du\, (u\bar u)^{N+1}
\bigl[ \mathcal{O}_{N}(z_{12}^ux)\bigr]_{l.t.}\,,
\label{eq:COPE}
\end{equation}
where $\varkappa_N=2(2N+3)/(N+1)!$. Here and below we use the following shorthand notation:
\begin{equation}
 \bar u =1-u\,,\quad z_{12}=z_1-z_2\,,\quad z^u_{12} = \bar u z_1 + u z_2\,.
\end{equation}
The conformal operator $\mathcal{O}_{N}$ is defined as
\begin{align}
\mathcal{O}_{N}(y)=&
(\partial_{z_1}\!+\!\partial_{z_2})^NC_N^{3/2}\left(
\frac{\partial_{z_1}\!-\!\partial_{z_2}}{\partial_{z_1}\!+\!\partial_{z_2}}\right)
\notag\\
&\times{O}(z_1x +y,z_2x+y)\Big|_{z_i=0},
\label{eq:On}
\end{align}
where $C_N^{3/2}(x)$ is the Gegenbauer polynomial.

We are able to find the contributions related to the leading-twist operator~(\ref{Olt})
in the time-ordered  product of two electromagnetic currents
\begin{equation}
T_{\mu\nu}=i\,T\Big\{j_\mu^{em}(x) j_\nu^{em}(0)\Big\}\,
\end{equation}
to the twist-four accuracy. The result has the form
\begin{align}\label{Tmn}
T_{\mu\nu} &= -\frac{1}{\pi^2x^4}\Big\{
x^{\alpha}\Big[S_{\mu\alpha\nu\beta} \mathbb{V}^\beta
+i\epsilon_{\mu\nu\alpha\beta} \mathbb{A}^\beta
\Big]
\notag\\
&+x^2\Big[(x_\mu\partial_\nu +x_\nu\partial_\mu) \mathbb{X}
+(x_\mu\partial_\nu-x_\nu\partial_\mu) \mathbb{Y}
\Big]
\Big\}\,,
\end{align}
where $\partial_\mu = \partial/\partial x^\mu$,
$S_{\mu\alpha\nu\beta}=g_{\mu\alpha}g_{\nu\beta}+g_{\nu\alpha}g_{\mu\beta}-g_{\mu\nu}g_{\alpha\beta}$
and the totally antisymmetric tensor is defined such that $\epsilon_{0123}=1$. The
expansion of $\mathbb{V}_\beta$ and $\mathbb{A}_\beta$ starts from  twist two,
$\mathbb{V}_\beta = \mathbb{V}^{t=2}_\beta + \mathbb{V}^{t=3}_\beta + \mathbb{V}^{t=4}_\beta+ \ldots$,
$\mathbb{A}_\beta = \mathbb{A}^{t=2}_\beta + \mathbb{A}^{t=3}_\beta + \mathbb{A}^{t=4}_\beta + \ldots$,
while $\mathbb{X} = \mathbb{X}^{t=4}+\ldots$ and $\mathbb{Y} = \mathbb{Y}^{t=4}+\ldots$
are twist-four.

It turns out that vector operators always appear to be antisymmetrized and axial-vector
symmetrized over the quark and antiquark positions, respectively, so we
define the corresponding combinations:
\begin{eqnarray}
{O}^{(-)}_{V}(z_1,z_2)&=\!&
\big[\bar q(z_1 x)\slashed{x}\, Q\,q(z_2 x )\big]_{l.t.} \!- (z_1\leftrightarrow z_2)\,,
\\
{O}^{(+)}_{A}(z_1,z_2)&=\!&
\big[\bar q(z_1 x)\slashed{x}\,\gamma_5\, Q\,q(z_2 x )\big]_{l.t.} \!+ (z_1\leftrightarrow z_2)\,.
\nonumber
\end{eqnarray}
The leading-twist expressions are well known and can be written as
(cf.~\cite{Balitsky:1987bk})
\begin{eqnarray}
\mathbb{V}^{t=2}_\mu
&=&\frac12 \partial_\mu
\int_0^1\!{du}\,{O}^{(-)}_{V}(u,0)\,,
\nonumber\\
\mathbb{A}^{t=2}_\mu
&=&\frac12\partial_\mu
\int_0^1\!{du}\,{O}^{(+)}_{A}(u,0)\,.
\end{eqnarray}
Note that separation of the leading-twist contributions $[\ldots]_{l.t.}$ from
the nonlocal operators by itself produces a series of kinematic power corrections
to the amplitudes, which are analogous to Nachtmann target mass corrections
to deep-inelastic scattering~\cite{Nachtmann:1973mr}. Such corrections
are discussed in detail 
in~\cite{Belitsky:2001hz,Belitsky:2000vx,Belitsky:2010jw,Geyer:2004bx,Blumlein:2006ia,Blumlein:2008di}.

For the twist-three functions we obtain
\begin{align}\label{VV3}
\mathbb{V}^{t=3}_\mu=&
\Big[i\mathbf{P}^\nu,\int_0^1 \!\!du\,\Big\{i\epsilon_{\mu\alpha\beta\nu}
x^\alpha \partial^\beta\widetilde{{O}}^{(+)}_{A}(u)
\notag\\&
+
\Big(S_{\mu\alpha\nu\beta}x^\alpha\partial^\beta+
\ln u \,\partial^\mu x^2\partial^\nu\Big)
\widetilde{{O}}^{(-)}_{V}(u)\Big\}\Big]\,,
\notag\\
\mathbb{A}^{t=3}_\mu=&
\Big[i\mathbf{P}^\nu,\int_0^1 \!\!du\,\Big\{i\epsilon_{\mu\alpha\beta\nu}
x^\alpha \partial^\beta\widetilde{{O}}^{(-)}_{V}(u)
\notag\\&
+
\Big(S_{\mu\alpha\nu\beta}x^\alpha\partial^\beta+
\ln u \,\partial^\mu x^2\partial^\nu\Big)
\widetilde{{O}}^{(+)}_{A}(u)\Big\}\Big]\,.
\end{align}
Here $\mathbf{P}_\nu$ is the momentum operator
\begin{align*}
 [i\mathbf{P}_{\!\!\nu}, q(y)]=\frac{\partial}{\partial y^\nu} q(y)\,,
&&
\langle p'|[\mathbf{P}_{\!\!\nu},O]|p\rangle = (p'-p)_\nu \langle p'|O|p\rangle
\end{align*}
and we used a notation
\begin{equation}\label{Ot}
\widetilde{{O}}^{(\pm)}_a(z)=\frac{1}{4}\int_{0}^{z}\! dw \,{O}^{(\pm)}_a(z,w)\,.
\end{equation}
One can easily check that $x^\mu \mathbb{V}_\mu^{t=3}=\partial^\mu \mathbb{V}_\mu^{t=3}=0$
and similarly $x^\mu \mathbb{A}_\mu^{t=3}=\partial^\mu \mathbb{A}_\mu^{t=3}=0$.
Note that the terms  in $\ln u $ in Eqs.~(\ref{VV3}) are themselves twist-four 
and can be dropped if the calculation is done to twist-three accuracy. 
The resulting simplified expression is in
agreement with Refs.~\cite{Radyushkin:2000ap,Belitsky:2000vx}. These terms must be
included, however, in order to ensure the separation of twist-three and twist-four
contributions.

The twist-four contributions $\mathbb{V}_\mu^{t=4}$, $\mathbb{A}_\mu^{t=4}$,
$\mathbb{X}^{t=4}$ and $\mathbb{Y}^{t=4}$ present our main result. 
In this case we did not find a simple nonlocal representation and write the answer in 
terms of integrals over the position of the local operators, cf. Eq.~(\ref{eq:COPE}). This form is equally 
well known and usually referred to as conformal OPE~\cite{Braun:2003rp}. 
It proves to be the most convenient for implementing 
the scale dependence in leading-twist GPDs~\cite{Mueller:2005ed,Kirch:2005tt}.
We obtain
\begin{widetext}
\begin{eqnarray}
\mathbb{V}^{t=4}_\mu&=&\frac12 \sum_{N,\text{odd}}\varkappa_N\frac1{(N+2)^2}\int_0^1 du \,(u\bar u)^{N+1}
\left\{
x_\mu\,[\widehat{\mathcal{O}}_N^V(u x)]_{l.t.}\,
+\frac12{N(N+3)} \int_0^1dv\,v^{N-1}\,  x^2\partial_\mu\,
[\widehat{\mathcal{O}}_N^V(uv x)]_{l.t.}\right\}\,,
\nonumber\\
\mathbb{A}^{t=4}_\mu&=&\frac14 x^2\partial_\mu \sum_{N,\text{even}}\varkappa_N\frac{N(N+3)}{(N+2)^2}
\int_0^1 du \,(u\bar u)^{N+1}\int_0^1dv\,v^{N-1}\,
[\widehat{\mathcal{O}}_N^A(uv x)]_{l.t.}\,,
\nonumber\\
\mathbb{X}^{t=4}&=&\frac14\sum_{N,\textrm{odd}} \varkappa_N\frac{N+1}{(N+2)^2}
\int_0^1 du\, (u\bar u)^{N}(u-\bar u)\int_0^1 dv\, v^{N-1}\, 
[\widehat{\mathcal{O}}_N^V(uv x)]_{l.t.}\,,
\nonumber\\
\mathbb{Y}^{t=4}&=&-\frac14\sum_{N,\textrm{odd}} \varkappa_N\frac{N+1}{(N+2)^2}
\int_0^1 du\, (u\bar u)^{N}(u^2+\bar u^2)\int_0^1 dv\, v^{N-1}\, 
[\widehat{\mathcal{O}}_N^V(uv x)]_{l.t.}\,.
\end{eqnarray}
\end{widetext}
Here $\widehat{\mathcal{O}}_N$ is defined as
the divergence of the leading-twist conformal
operator, cf. $\mathcal{O}_2$ in~Eq.\,(\ref{eq:O1O2}):
\begin{eqnarray}
\widehat{\mathcal{O}}_N(y)&=&\frac1{N+1}\frac{\partial}{\partial x^\mu} 
\bigl[i\mathbf{P}^\mu,\mathcal{O}_N(y)\bigr]
\notag\\&=&
\bigl[i\mathbf{P}^\mu,\mathcal{O}_{\mu\mu_1\ldots\mu_N}(y)\bigr] x^{\mu_1}\ldots x^{\mu_N}\,.
\end{eqnarray}
One of the two integrals over the position of $\widehat{\mathcal{O}}_N$ can easily 
be taken, resulting in slightly more lengthy expressions.
 
Note that the operator $\mathcal{O}_1$ in Eq.\,(\ref{eq:O1O2}),  which
reads $[i\mathbf{P}_\mu[i\mathbf{P}^\mu, \mathcal{O}_N]$ in our present
notation, does not contribute to the answer
for our special choice of the correlation function $T\{j_\mu(x)j_\nu(0)\}$.
The T-product with generic positions of the currents, $T\{j_\mu(z_1x)j_\nu(z_2x)\}$, 
includes both operators. The corresponding result is much more cumbersome 
and will be given elsewhere. 

For comparison we rewrite the leading-twist 
contribution in the same form:
\begin{equation}
 \mathbb{V}^{t=2}_\mu = 
 \partial_\mu \sum_{N,\mathrm{odd}}\frac{\varkappa_N}{N+2}
\int_0^1\!du\, u^N \bar u^{N+2}\, \mathcal{O}^V_{N}(ux)\,,
\end{equation} 
where $\mathcal{O}^V_{N}(ux)$ is the conformal operator
(\ref{eq:On}) at the space-time position $ux$.

Conservation of the electromagnetic current implies that 
\begin{align}\label{cc}
\partial^\mu T_{\mu\nu}(x)=0\,, &&\partial^\nu T_{\mu\nu}(x)=i[\mathbf{P}^\nu, T_{\mu\nu}(x)]\,.
\end{align}
We have checked that these identities are satisfied up to twist-5 terms.

For completeness we give the expression for the operator $[i\mathbf{P}_\mu, \partial^\mu O(z_1,z_2)]$
entering the twist-three functions $\mathbb{V}^{t-3}_\mu$, $\mathbb{A}^{t-3}_\mu$
in terms of $\widehat{\mathcal{O}}_N$:
\begin{eqnarray}
&&[i\mathbf{P}_{\!\mu}, \partial^\mu O(z_1,z_2)]=
\frac12 S^+ \!\! \int_0^1 \!\!\!\!u du\, [i\mathbf{P}_\mu[i\mathbf{P}^\mu\!, O(uz_1,uz_2)]]
\nonumber\\&&+
\sum_{N}\!\varkappa_N(N\!+\!1)^2 z_{12}^N \!\int_0^1 \!\!\!dv\, v^{N}
\!\!\int_0^1 \!\!\!du\, (u\bar u)^{N+1} \widehat{\mathcal{O}}_N(v z_{12}^u x),
\nonumber\\
\end{eqnarray}
where $S^+=z_1^2\partial_{z_1}+z_2^2\partial_{z_2}+2z_1+2z_2$. In phenomenological applications it 
can be advantageous to use relations of this kind to rewrite all contributions of 
$\widehat{\mathcal{O}}_N$ in terms of $[i\mathbf{P}_\mu, \partial^\mu  O(z_1,z_2)]$.

To summarize, we have given a complete expression for the time-ordered product 
of the two electromagnetic currents that resums all kinematic corrections 
to the twist-four accuracy. The results have immediate applications to the 
studies of deeply-virtual Compton scattering and $\gamma^*\to (\pi,\eta,\ldots) +\gamma$
transition form factors. The twist-four terms
calculated in this work give rise {\em both} to a $\sim t/Q^2$ correction and 
the target mass correction $\sim m^2/Q^2$ for DVCS, 
whereas for the transition form factors these two effects
are indistinguishable as there is only one mass scale.
We remark that the distinction between the kinematic  corrections 
due to contributions of 
leading-twist~\cite{Radyushkin:2000ap,Belitsky:2001hz,Belitsky:2000vx,Belitsky:2010jw,%
Geyer:2004bx,Blumlein:2006ia,Blumlein:2008di} and higher-twist operators considered in our work
is not invariant under translations along the line connecting the currents and has no physical 
meaning.  Such corrections must always be summed up.
Concrete applications go beyond the tasks of this letter.
   
{\bf Acknowledgements}
The work by A.M. was supported by the DFG, grant BR2021/5-2,
and RFFI, grant 09-01-93108.


\end{document}